\documentclass[11pt,a4paper]{article}

\usepackage[T1]{fontenc}
\usepackage[utf8]{inputenc}
\usepackage{lmodern}
\usepackage[margin=2.7cm]{geometry}
\usepackage{microtype}
\usepackage{graphicx}
\usepackage{amsmath,amsfonts,amssymb,amsthm}
\usepackage{bm}
\usepackage{multirow}
\usepackage{booktabs}
\usepackage{tabularx}
\usepackage{authblk}
\usepackage[round,authoryear]{natbib}
\usepackage[dvipsnames]{xcolor}
\usepackage[colorlinks=true,linkcolor=MidnightBlue,citecolor=MidnightBlue,urlcolor=MidnightBlue]{hyperref}
\usepackage[capitalize,noabbrev]{cleveref}

\newcommand{\spaceY}{\mathcal{Y}}
\newcommand{\spaceX}{\mathcal{X}}
\newcommand{\nsetR}{\mathbb{R}}

\newcommand{\prob}[1]{\mathbb{P}\left(#1\right)}

\theoremstyle{definition}
\newtheorem{definition}{Definition}

\theoremstyle{plain}
\newtheorem{theorem}{Theorem}

\theoremstyle{remark}

\begin{document}

\title{Practical Advantages of Online Conformal Prediction for Data Streams}
\author[1]{Stéphane Chrétien}
\author[2]{Maryam Hoseini-Behbahani}
\author[1]{Guillaume Metzler}
\affil[1]{Université Lumière Lyon 2 and ERIC Laboratory, Université Lyon 1 and Université Lyon 2, 5 avenue Pierre Mendès-France, 69500 Bron, France\\
\texttt{stephane.chretien@univ-lyon2.fr}, \texttt{guillaume.metzler@univ-lyon2.fr}}
\affil[2]{Université Lyon 1, France\\
\texttt{maryam.hoseini-behbahani@etu.univ-lyon1.fr}. Work conducted during an internship at ERIC Laboratory.}
\date{}
\maketitle

\begin{abstract}
Full conformal prediction offers finite-sample marginal coverage guarantees under
exchangeability, but exact computation can be costly, while its classical guarantees
do not directly extend to non-stationary data streams. We compare four sequential
conformal regression procedures with a weighted full-conformal benchmark in terms
of empirical coverage, prediction interval width, and computational cost. The
procedures combine ordinary or forgetting-factor least squares with uniform or
exponentially weighted empirical residual quantiles, or with online gradient-based
quantile tracking. Experiments cover an independent and identically distributed
control, two synthetic distribution-shift scenarios, and four real-world data
streams. Adapting the regression model as well as the conformal threshold
substantially improves the coverage--width tradeoff under drift. The
quantile-tracking procedure requires $O(p^2)$ time and memory per observation,
independently of stream length, although its coverage can deteriorate after
abrupt changes. These results illustrate the practical tradeoffs between
predictor adaptation, uncertainty quantification, and computational resources.
\end{abstract}

\par

\section{Introduction}
\label{sec:intro}

The deployment of machine learning models in safety-critical and resource-constrained environments, energy markets, finance, sensor networks, has made uncertainty quantification a first-class requirement.
A point prediction, however accurate on average, is of limited use without a well-calibrated measure of confidence. 
Conformal Prediction (CP)~\citep{vovk2005algorithmic} instead gives \emph{finite-sample}, \emph{distribution-free} marginal coverage under the sole assumption of exchangeability, wrapping any pre-trained score function into a prediction set $C_\alpha(X)$ satisfying $\prob{Y\in C_\alpha(X)}\ge 1-\alpha$ for any user-chosen $\alpha$. 
Full Conformal Prediction (FCP) achieves the tightest such set, but at a prohibitive cost: for every test point and every candidate label $y\in\spaceY$, the model must be re-scored or re-trained on the augmented sample. 
Split CP~\citep{lei2018distribution} avoids this by fixing the model and calibrating on a held-out set, trading statistical efficiency (only part of the data drives the quantile) for a single forward pass. 
Neither FCP nor Split CP is designed for non-stationary streams, where the joint distribution of $(X,Y)$ drifts over time, a setting that is the rule rather than the exception in the applications above.

Several lines of work address non-stationarity within CP:  
Adaptive Conformal Inference (ACI)~\citep{gibbs2021adaptive} updates the coverage level online via a proportional controller on the pinball loss, recovering long-run coverage under drift; an aggregated variant improves robustness further.
Weighted CP~\citep{barber2023conformal,tibshirani2019conformal} down-weights distant observations via importance ratios, giving a principled oracle for non-exchangeable streams. 
Online quantile-tracking methods with regret guarantees~\citep{angelopoulos2024online} bound a scalar tracker directly.

None of these jointly resolves two distinct requirements: \textit{(i)} valid coverage and tight width under drift, and \textit{(ii)} per-step cost independent of the stream length $t$. 
ACI-style methods satisfy \textit{(ii)} trivially (a scalar update) but not \textit{(i)}: they leave the predictor untouched, so once the base model goes stale the quantile update alone must absorb an error it did not cause. 
Weighted CP satisfies \textit{(i)} but not \textit{(ii)}: its exact set is recomputed from the full re-weighted history at every step, which is exactly the bottleneck that made FCP impractical to begin with. 
Regret-bounded quantile trackers satisfy \textit{(ii)} but treat the underlying model as given, and, to our knowledge, no prior work benchmarks the resulting pipeline against the exact full/weighted-CP set it implicitly approximates, in either regression or classification.

In the present work, we empirically investigate whether, under distribution shift, an online variant of conformal prediction can match full conformal prediction (full CP) at a fraction of its cost. 
The studied method, QT+WLS, pairs a forgetting-factor model update (Recursive Least Squares, RLS) with an online-gradient quantile tracker, so that its memory and per-step cost are $O(p^2)$, where $p$ is the dimension of the feature space, and independent of $t$, the iteration of the online setting, unlike weighted or exponentially weighted variants that still carry a growing residual array. 
We compare four online conformal regression methods against a full CP oracle using the entire observed history (an exact weighted homotopy solver, not a windowed approximation) on three synthetic regimes and the real world ELEC2 dataset.

\section{Background}
\label{sec:background}

In this section, we introduce the conformal settings considered in the paper: \textit{(i)} full conformal prediction, \textit{(ii)} weighted conformal prediction, and \textit{(iii)} online conformal prediction, before providing further details on the exact algorithms studied in \Cref{sec:methods}.

Let $\spaceX\subseteq\nsetR^p$ be the input space and $\spaceY$ the output space ($\spaceY\subseteq\nsetR$ for regression, a finite label set for classification). 
We observe a stream $(X_1,Y_1),(X_2,Y_2),\ldots$ indexed by $t=1,2,\ldots$. 
For a training horizon $T_{\rm tr}$ and calibration horizon $T_{\rm cal}$, let $T_0=T_{\rm tr}+T_{\rm cal}$ denote the start of the online phase.

\begin{definition}[Exchangeability]
\label{def:exchangeability}
A sequence $Z_1,\ldots,Z_n$ is \emph{exchangeable} if its joint distribution is invariant under any permutation of the indices. 
Every \textit{i.i.d.} sequence is exchangeable; the converse does not hold in general.
\end{definition}

\begin{definition}[Nonconformity score]
\label{def:score}
A \emph{nonconformity measure} is $s:\spaceX\times\spaceY\to\nsetR$, large when $(x,y)$ is atypical relative to the training data. 
For regression we use the absolute residual $s_t=|Y_t-\hat Y_t|$, where $\hat Y_t$ is formed \emph{before} observing $Y_t$.
\end{definition}

\subsection{Full Conformal Prediction}
\label{subsec:full_cp}

We briefly remind the full conformal setting, its associated complexity and its coverage guarantees.

\begin{definition}[Full CP set]
\label{def:full_cp}
Given $S_n=\{(X_i,Y_i)\}_{i=1}^n$, a test point $x_{n+1}$, a candidate
label $y\in\spaceY$, and score $s$, let $S_n^y=S_n\cup\{(x_{n+1},y)\}$,
$h^y$ the model trained on $S_n^y$, and $s_i^y=s(h^y(X_i),Y_i)$ for
$i\le n$, $s_{n+1}^y=s(h^y(x_{n+1}),y)$. Define
\begin{equation}
  \hat{q}^y = \operatorname{quantile}_{1-\alpha}\!\left(s_1^y,\ldots,s_n^y,
    \tfrac{\lceil(n+1)(1-\alpha)\rceil}{n}\right),
\end{equation}
and $C_\alpha^{\rm full}(x_{n+1}) = \{y\in\spaceY : s_{n+1}^y \le
\hat{q}^y\}$.
\end{definition}

\begin{theorem}[Marginal coverage of full CP {\citep{vovk2005algorithmic}}]
\label{thm:full_cp}
If $(X_1,Y_1),\ldots,(X_n,Y_n),(x_{n+1},Y_{n+1})$ are exchangeable, then
for any $\alpha\in(0,1)$,
\begin{equation}
  1-\alpha \le \prob{Y_{n+1}\in C_\alpha^{\rm full}(x_{n+1})} \le
  1-\alpha+\tfrac{1}{n+1}.
\end{equation}
\end{theorem}

\subsection{Online CP}
\label{subsec:online_cp}

Conformal prediction was originally introduced in an online setting, where
examples arrive one at a time and the set for $Y_t$ is built from the
first $t-1$ observations \citep{vovk2002online, vovk2005algorithmic}.
Online conformal prediction (OCP) keeps this sequential protocol but
drops the exchangeability assumption, so as to handle streams whose
distribution drifts. At each $t \ge T_0$, the learner observes $X_t$,
outputs a set $\hat C_t(X_t)$ \emph{before} $Y_t$ is revealed, then
observes $Y_t$ and updates its internal state. This state has two
components: the base model, which produces the prediction $\hat Y_t$ and
hence the score $R_t$, and the conformal threshold $\hat q_t$, which
turns past scores into a prediction set.

The existing literature on OCP under distribution shift essentially focuses on the threshold. Adaptive Conformal Inference (ACI) \citep{gibbs2021adaptive, gibbs2024conformal} adjusts the working miscoverage level from the observed coverage errors and guarantees long-run coverage under arbitrary shifts, and several works refine the choice of its step size, through aggregation \citep{zaffran2022adaptive}, strongly adaptive online learning \citep{bhatnagar2023improved}, or decaying step sizes with regret bounds \citep{angelopoulos2024online}.
A complementary view interprets the threshold update as a feedback controller \citep{angelopoulos2023pid}. 
In most of these works, the base model is treated as given. 
The rules used to update the model and the threshold, and how they compare to the full-CP oracle, are the object of the next section.

\section{Online Conformal Methods}
\label{sec:methods}

All methods follow the same three-phase protocol: \textit{(i)} training
on the first $T_{\rm tr}$ observations; \textit{(ii)} calibration on the
next $T_{\rm cal}$, collecting honest out-of-sample scores (predictions
made before the model updates); \textit{(iii)} online prediction from
$T_0$ onward, where methods differ in how the model and quantile adapt.
\Cref{tab:complexity} summarizes every method compared in this paper for regression with their associated complexity.

\begin{table}[t]
\centering
\caption{Per-step cost and memory for each of the studied methods where $p$ stands for the dimension of the feature space and $t$ the iteration.}
\label{tab:complexity}
\begin{tabular}{lcc}
\hline
Method & Time per step & Memory \\
\hline
Full CP (oracle)    & $O(p^3+t\log t)$ & $O(t\,p)$ \\
CP+LS               & $O(p^2+t)$ & $O(p^2+t)$ \\
nexCP+LS, nexCP+WLS & $O(p^2+t\log t)$ & $O(p^2+t)$ \\
QT+WLS              & $O(p^2)$ & $O(p^2)$ \\
\hline
\end{tabular}
\end{table}

\subsection{Quantile Updates}
\label{subsec:quantile_updates}

Let $R_i=|Y_i-\hat Y_i|$,
and $\mathcal{R}_t=\{R_i\}_{i<t}$.

\paragraph{Uniform weights.} $\hat q_{t,{\rm uni}} =
\mathrm{quantile}_{1-\alpha}(\mathcal{R}_t)$: appropriate under
stationarity, slow to adapt after a shift since all past residuals carry
equal weight.

\paragraph{Exponential weights.} $w_{i,t}=\rho^{t-i}$, $\rho\in(0,1)$,
giving the weighted empirical quantile
\begin{align}
    \hat q_{t,{\rm exp}}=\inf\{r:\sum_{i:R_i\le r}w_i/\sum_j w_j \ge
1-\alpha\}
\end{align}~\citep{barber2023conformal}. The effective sample size
$N_{\rm eff}^\rho \simeq 1/(1-\rho)$ for large $t$ represents the number
of influential past examples, i.e., how quickly old observations are
forgotten. This still requires storing and re-sorting a growing residual
array, $O(t\log t)$ per step.

\paragraph{OGD quantile tracking.} Inspired by ACI~\citep{gibbs2021adaptive}
and the quantile-tracking view of the conformal
P-controller~\citep{angelopoulos2023conformal}, we maintain a single
scalar threshold $\theta_t$ via online subgradient descent on the
pinball loss:
\begin{equation}
    \theta_{t+1} = \max\!\left(0,\ \theta_t + \eta
    \begin{cases}
        (1-\alpha) & R_t>\theta_t,\\
        -\alpha    & R_t\le\theta_t,
    \end{cases}\right)
    \label{eq:ogd}
\end{equation}
with step size $\eta>0$: constant memory and computation per step,
regardless of $t$, and no residual array at all. Its validity follows
from the online ACI guarantee~\citep{gibbs2021adaptive} rather than from
convergence to an FCP limit.

\subsection{Model definitions and updates}
\label{subsec:model_updates}

\paragraph{Time varying linear model}
At each time $t$, the learner observes a covariate vector
$x_t\in\nsetR^p$ and predicts a scalar response $y_t\in\nsetR$
(\Cref{tab:covariates}). The integer $p$ is the \emph{order} of the
model, i.e.\ its number of coefficients; when an intercept is used, it
is one of the $p$ columns of $x_t$ (a constant column).
All methods share the same prediction function, which is linear in the
covariates:
\begin{equation}
  \hat y_t = x_t^\top \hat\beta_t ,
  \qquad \hat\beta_t\in\nsetR^p ,
  \label{eq:prediction}
\end{equation}
where $\hat\beta_t$ is estimated from the observations $i<t$ only, so
that $\hat y_t$ is formed before $y_t$ is revealed. The data are
modeled as
\begin{equation}
  Y_t = x_t^\top\beta_t + \varepsilon_t ,
  \label{eq:truemodel}
\end{equation}
with a possibly time-varying coefficient vector $\beta_t$. For the
synthetic streams, \eqref{eq:truemodel} holds exactly with
$\varepsilon_t\sim\mathcal N(0,1)$ and a known $\beta_t$. 

\paragraph{Least-squares model (CP+LS, nexCP+LS).}
The static model uses all past observations with equal weight:
\begin{align}
    \hat\beta_t^{\rm LS}
  = \arg\min_{\beta\in\nsetR^p}\sum_{i<t}\bigl(y_i-x_i^\top\beta\bigr)^2
  = A_t^{-1}c_t ,\\
  A_t=\sum_{i<t}x_ix_i^\top,\;
  c_t=\sum_{i<t}x_iy_i .
  \label{eq:ols}
\end{align}
It is initialized on the first $T_{\rm tr}$ observations (which requires
$A_{T_{\rm tr}+1}$ to be invertible) and keeps absorbing every new
observation.

\paragraph{Forgetting model (nexCP+WLS, QT+WLS).}
The adaptive model discounts old observations exponentially with the
forgetting factor $\rho_{\rm LS}\in(0,1)$:
\begin{align}
    \hat\beta_t^{\rm WLS}
  = \arg\min_{\beta\in\nsetR^p}\sum_{i<t}\rho_{\rm LS}^{\,t-1-i}
    \bigl(y_i-x_i^\top\beta\bigr)^2
  = S_t^{-1}b_t ,\\
  S_t=\sum_{i<t}\rho_{\rm LS}^{\,t-1-i}x_ix_i^\top,\;
  b_t=\sum_{i<t}\rho_{\rm LS}^{\,t-1-i}x_iy_i .
  \label{eq:wls}
\end{align}
Its effective memory is $N_{\rm eff}=1/(1-\rho_{\rm LS})\approx111$
observations.

\paragraph{Scores and prediction sets.}
Whichever model is used, $\hat y_t$ in \eqref{eq:prediction} yields the
score $R_t=|y_t-\hat y_t|$ and the interval
\begin{equation}
  \hat C_t(x_t)=\bigl[\hat y_t-\hat q_t,\ \hat y_t+\hat q_t\bigr],
  \qquad \text{width}=2\hat q_t ,
\end{equation}
where $\hat q_t$ is the uniform, exponentially weighted or OGD quantile
of \Cref{subsec:quantile_updates}. The methods thus differ only through
$\hat\beta_t$ and $\hat q_t$.

\paragraph{Oracle model.}
For a candidate value $y$ of the response, the full-CP oracle refits the
weighted model on the augmented sample
$\{(x_i,y_i)\}_{i\le t}\cup\{(x_t,y)\}$, with the same weights
$\tilde w_i=\rho_{\rm LS}^{\,t+1-i}$:
\begin{equation}
  \hat\beta^{\,y}
  = \arg\min_{\beta}\sum_{i\le t+1}\tilde w_i\bigl(y_i-x_i^\top\beta\bigr)^2 ,
  \quad (x_{t+1},y_{t+1})=(x_t,y),
\end{equation}
so that the oracle and nexCP+WLS fit the same weighted problem.

\begin{table}[t]
\centering
\caption{Covariates, response and model order $p$ of each stream. ``1''
denotes the constant column of the intercept, which counts in $p$.
Temperatures are in $^\circ$C, humidities in \%, pressures in
mm\,Hg (Energy) or hPa (PRSA), and wind speeds in m/s; the Bike
weather covariates are normalized to $[0,1]$.}
\label{tab:covariates}
\footnotesize
\setlength{\tabcolsep}{3pt}
\begin{tabularx}{\textwidth}{@{}lXlc@{}}
\toprule
Stream & Covariates $x_t$ & Response $y_t$ & $p$ \\
\midrule
Synthetic & four independent $\mathcal N(0,1)$ variables & $x_t^\top\beta_t+\varepsilon_t$ & 4 \\
ELEC2 & electricity price and demand in New South Wales and in Victoria & electricity transfer & 4 \\
Energy & energy use of lights, kitchen temperature and humidity, outdoor temperature, humidity, pressure and wind speed, 1 & energy use of appliances (Wh) & 8 \\ % CHECK
Bike & temperature, humidity, wind speed, working-day indicator, 1 & number of rented bikes & 5 \\ % CHECK
PRSA & dew point, temperature, pressure, cumulated wind speed, 1 & PM2.5 concentration & 5 \\ % CHECK
\bottomrule
\end{tabularx}
\end{table}

\paragraph{OLS updates via Sherman--Morrison.} For the two methods
using a static (unweighted) least-squares regressor, we avoid
re-inverting a $p\times p$ matrix at every step via the Sherman--Morrison
rank-one update, at $O(p^2)$ per step with no loss of exactness.

\paragraph{RLS updates.} For the two methods that must also let the base model forget outdated observations, we use Recursive Least Squares with forgetting factor $\rho_{\rm LS}$, also $O(p^2)$ per step; observations older than roughly $N_{\rm eff}=1/(1-\rho_{\rm LS})$ steps are effectively forgotten.

The main empirical distinction (\Cref{subsec:res}) is between methods that adapt only the quantile (CP+LS, nexCP+LS) and those that also adapt the model via forgetting (nexCP+WLS, QT+WLS): model adaptation turns out to be necessary, not optional, under drift.

\section{Experiments and Results}
\label{sec:expres}

\subsection{Results and Interpretation}
\label{subsec:res}

\begin{figure}
    \centering \includegraphics[width=0.45\linewidth]{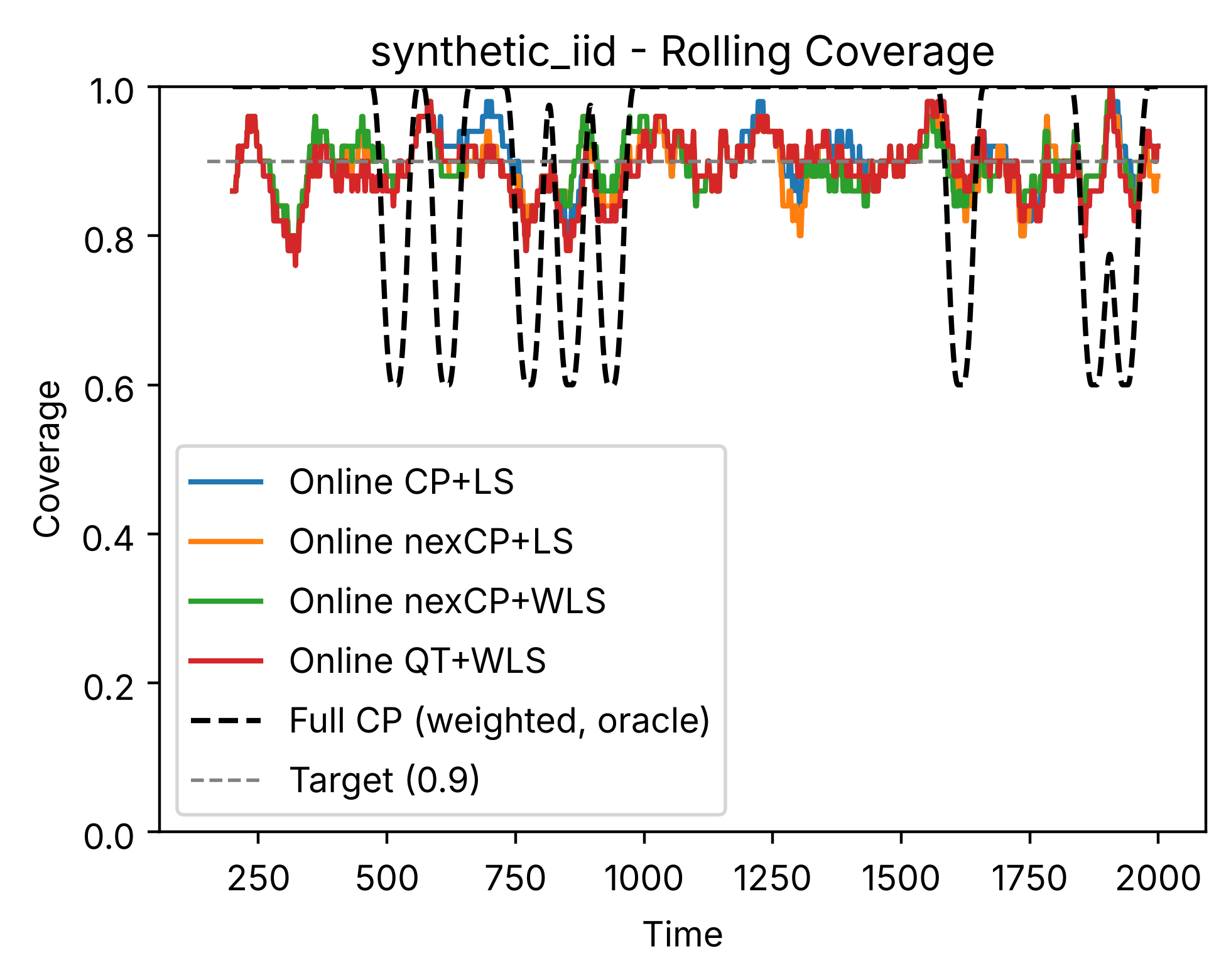} \includegraphics[width=0.45\linewidth]{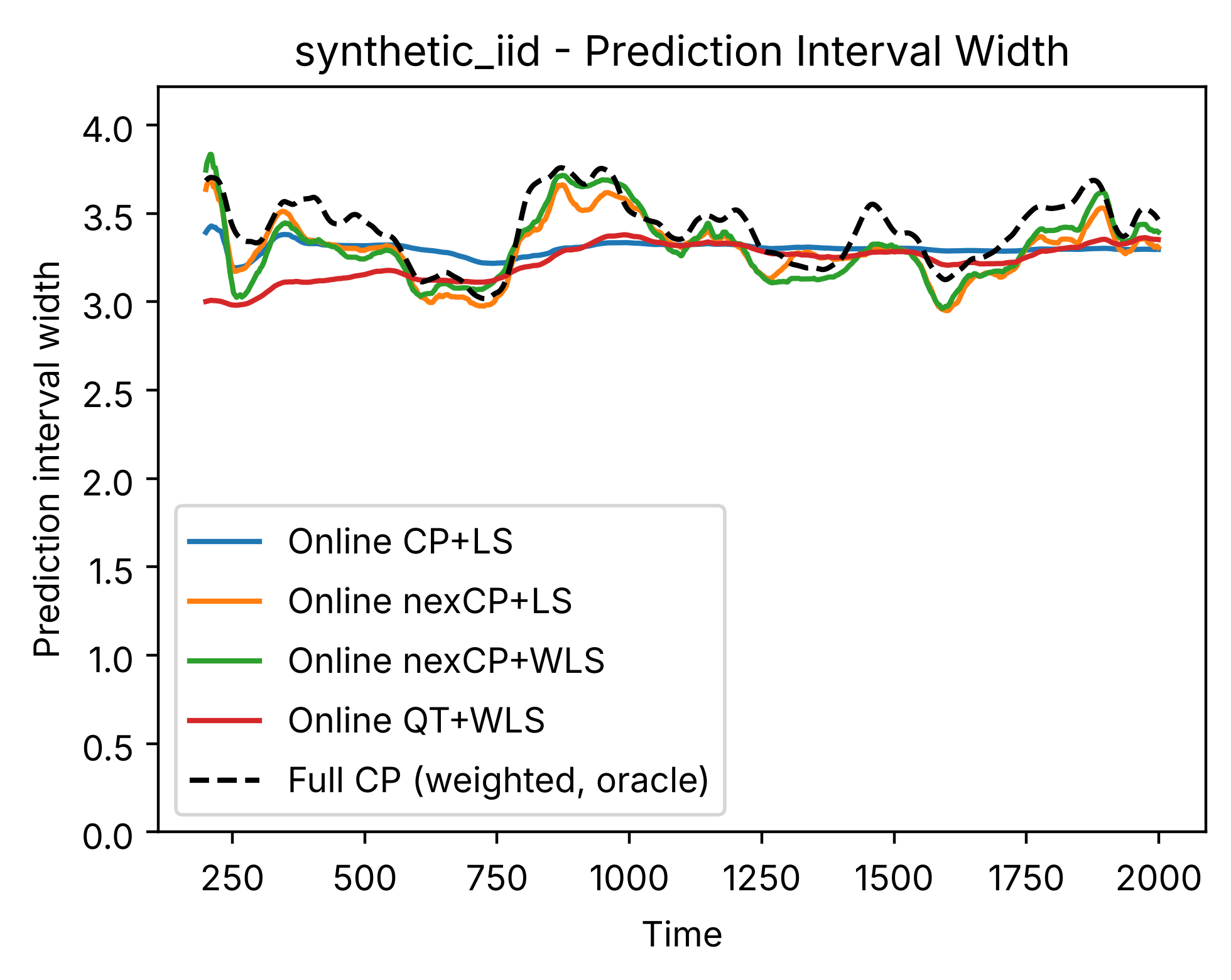}
    \includegraphics[width=0.45\linewidth]{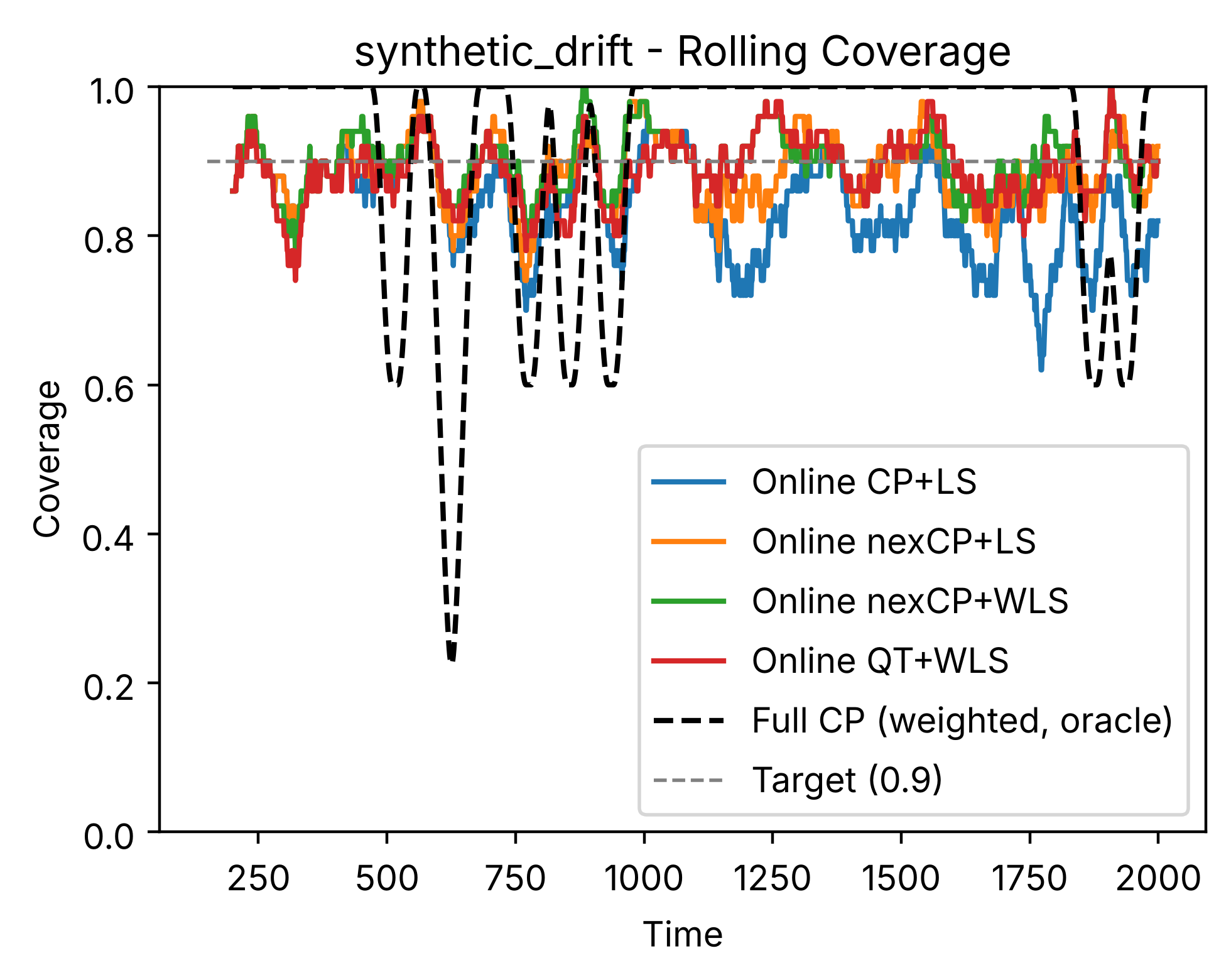} \includegraphics[width=0.45\linewidth]{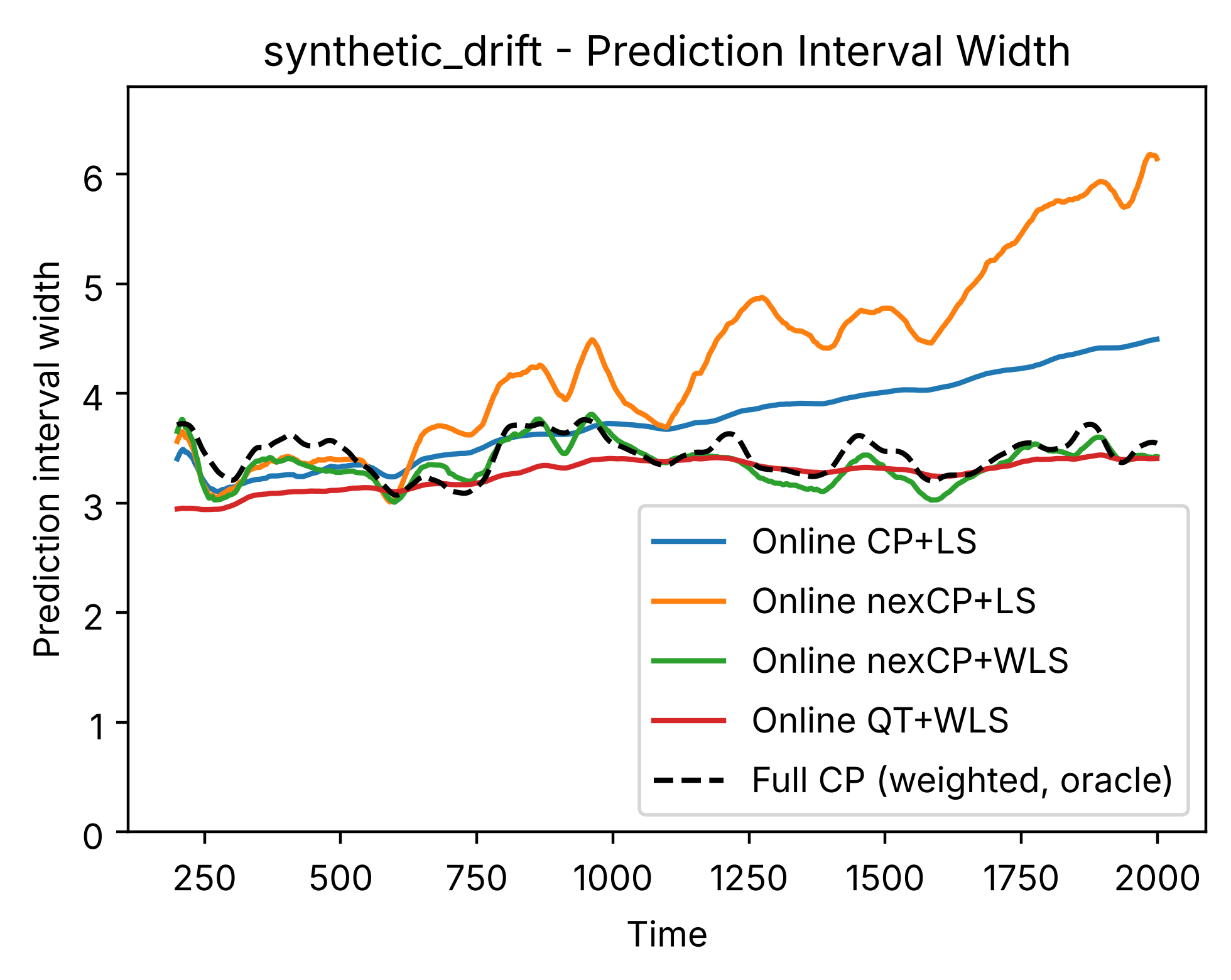}
    \includegraphics[width=0.45\linewidth]{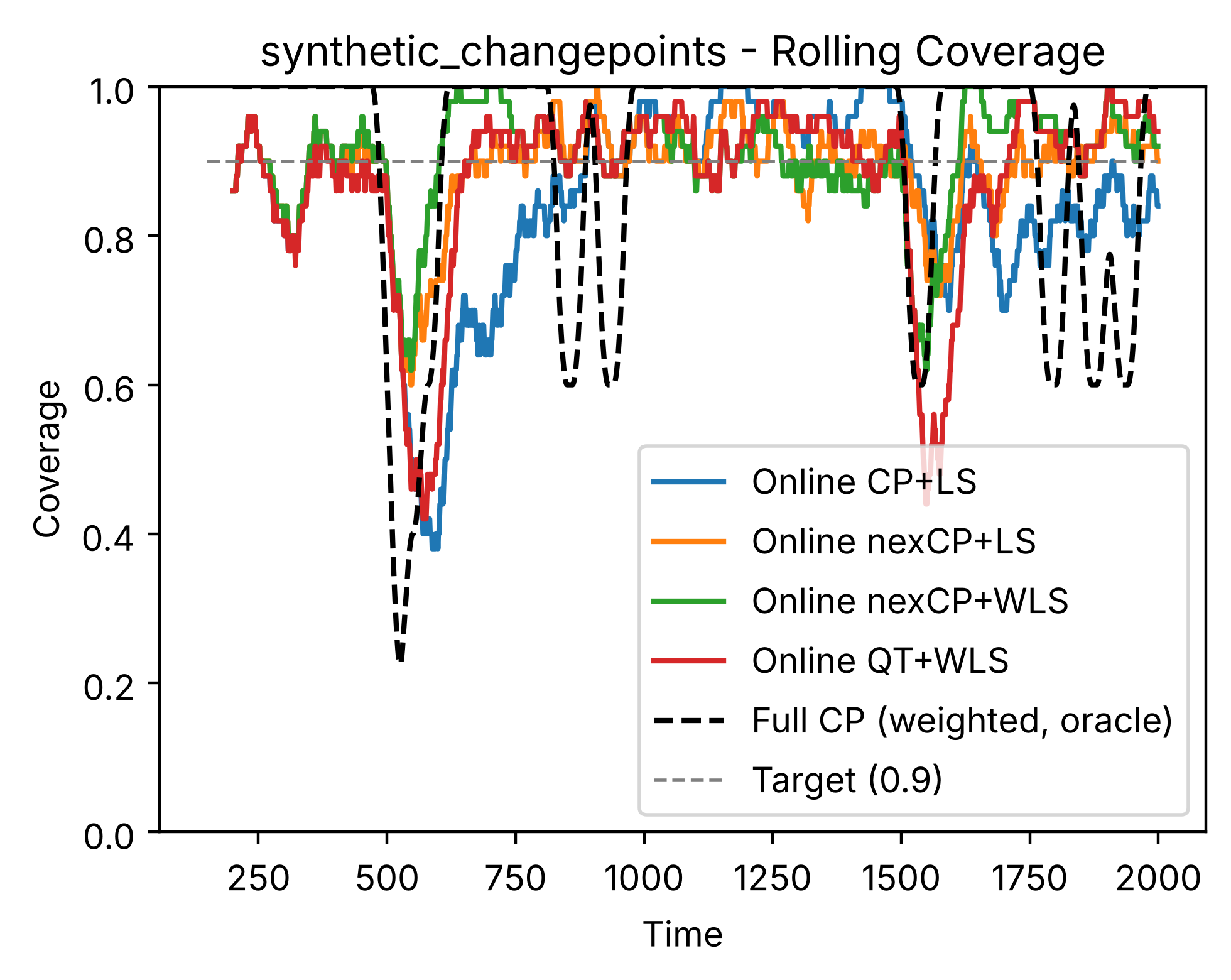} \includegraphics[width=0.45\linewidth]{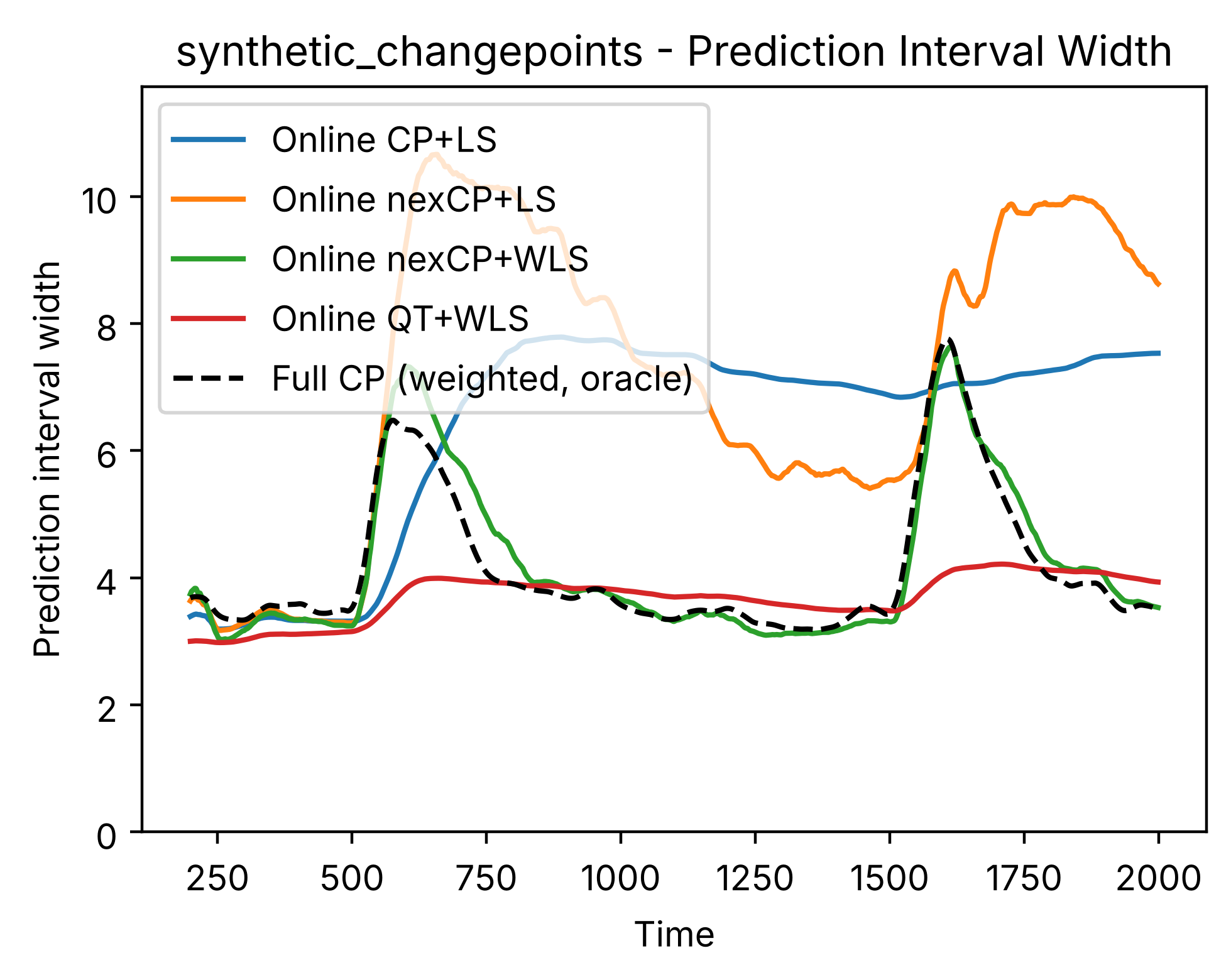}
    \includegraphics[width=0.45\linewidth]{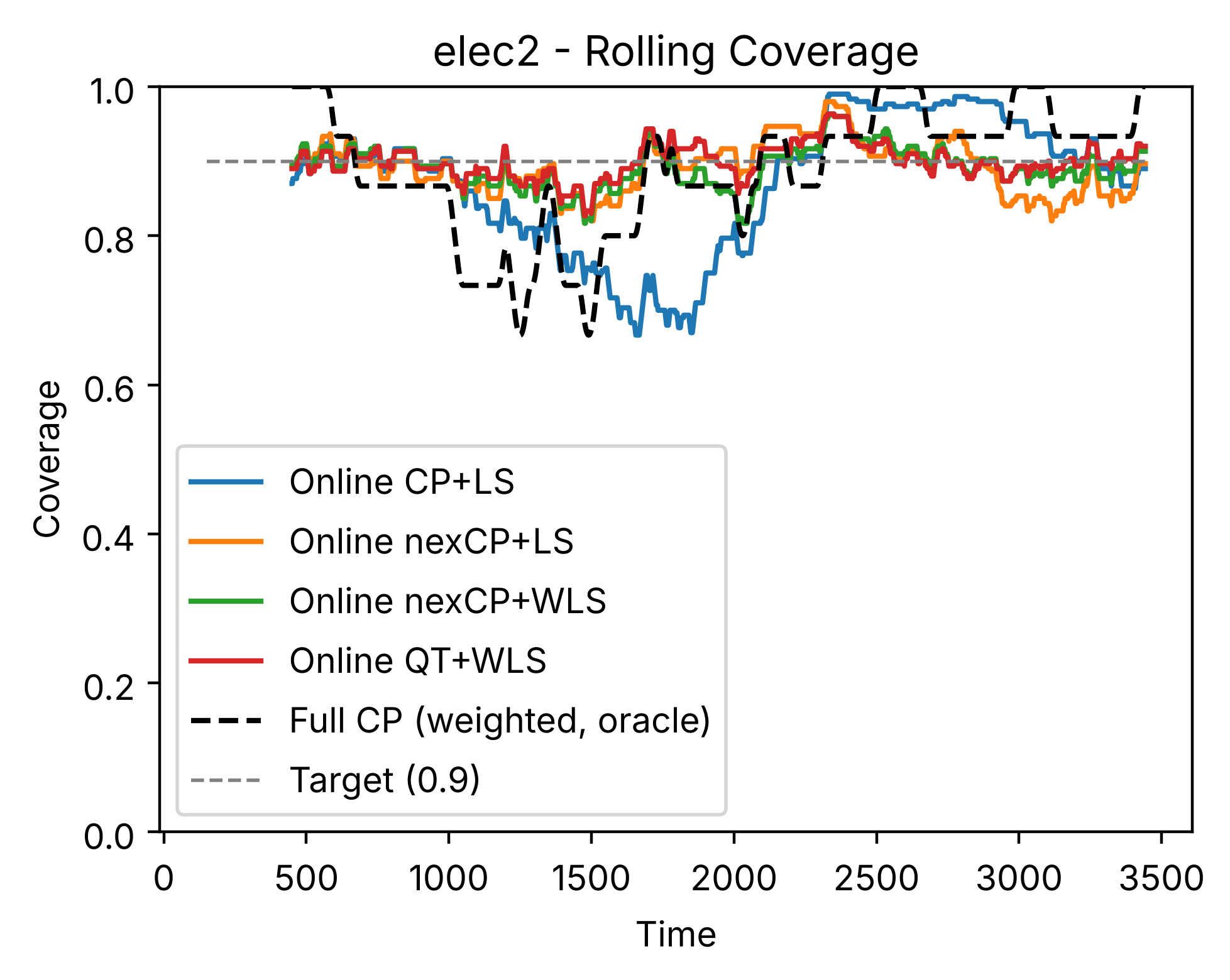} \includegraphics[width=0.45\linewidth]{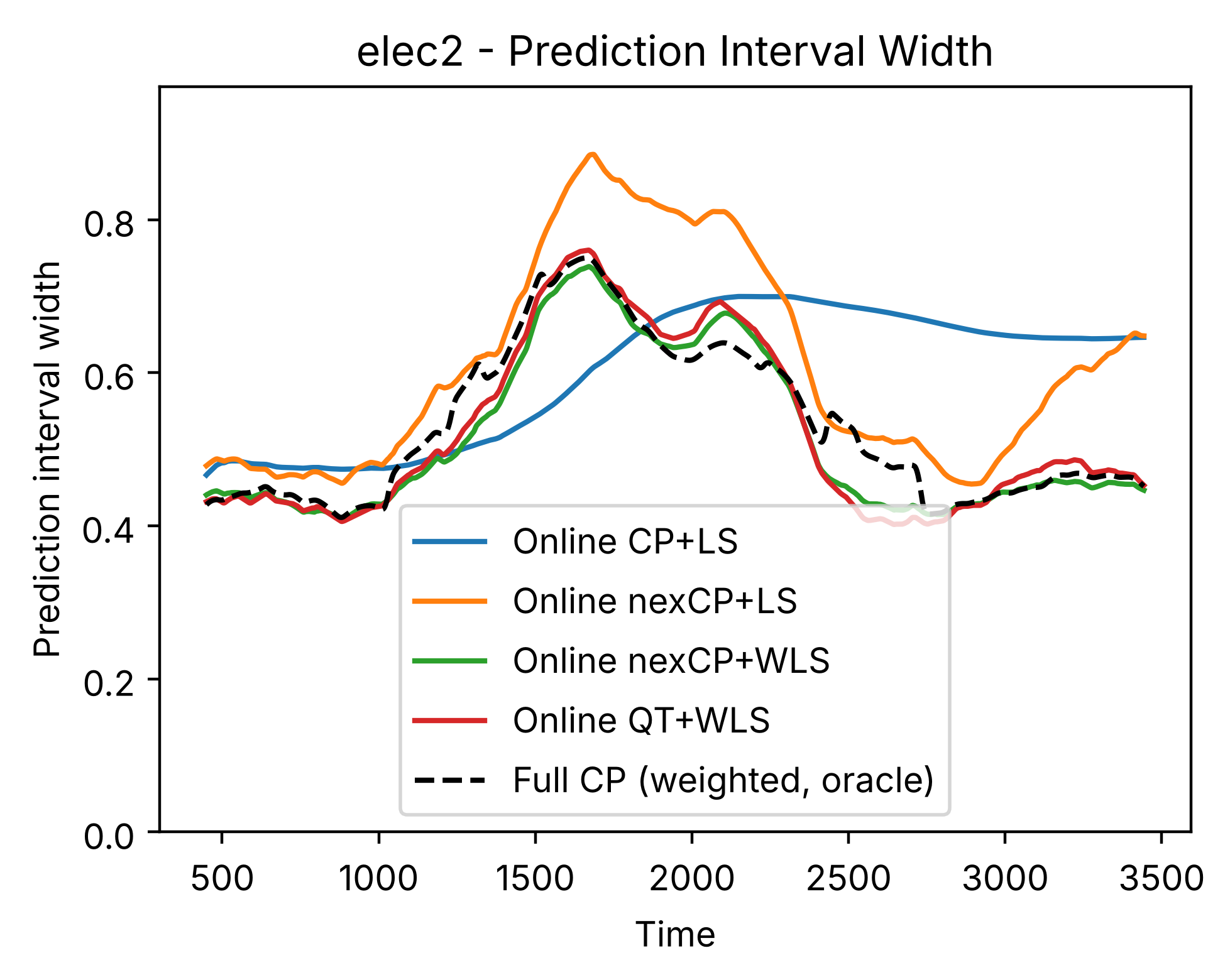}
    \caption{Rolling coverage (left) and interval width (right), online methods vs. the full-CP oracle on the three synthetics and the Elec2 datasets.}
    \label{fig:cov-width}
\end{figure}

\begin{table}[t]
\centering
\caption{Average coverage (target $0.90$) and width over the online phase.
Synthetic streams: mean $\pm$ standard deviation over 20 realizations.}
\label{tab:results}
\setlength{\tabcolsep}{3pt}
\begin{tabular}{llcc}
\toprule
Stream & Method & Coverage & Width \\
\midrule
\multirow{5}{*}{i.i.d.} & Full CP (oracle) & .912 $\pm$ .021 & 3.41 $\pm$ .05 \\
 & CP+LS & .897 $\pm$ .006 & 3.28 $\pm$ .07 \\
 & nexCP+LS & .896 $\pm$ .002 & 3.30 $\pm$ .05 \\
 & nexCP+WLS & .897 $\pm$ .003 & 3.31 $\pm$ .05 \\
 & QT+WLS & .894 $\pm$ .010 & \textbf{3.27 $\pm$ .10} \\
\midrule
\multirow{5}{*}{Changepoints} & Full CP (oracle) & .911 $\pm$ .028 & 4.15 $\pm$ .06 \\
 & CP+LS & .844 $\pm$ .007 & 6.26 $\pm$ .13 \\
 & nexCP+LS & .891 $\pm$ .003 & 7.13 $\pm$ .14 \\
 & nexCP+WLS & .903 $\pm$ .004 & 4.20 $\pm$ .08 \\
 & QT+WLS & .876 $\pm$ .009 & \textbf{3.78 $\pm$ .11} \\
\midrule
\multirow{5}{*}{Drift} & Full CP (oracle) & .912 $\pm$ .024 & 3.45 $\pm$ .05 \\
 & CP+LS & .836 $\pm$ .008 & 3.73 $\pm$ .06 \\
 & nexCP+LS & .886 $\pm$ .003 & 4.30 $\pm$ .09 \\
 & nexCP+WLS & .898 $\pm$ .002 & 3.36 $\pm$ .05 \\
 & QT+WLS & .892 $\pm$ .010 & \textbf{3.30 $\pm$ .09} \\
\midrule
\multirow{5}{*}{ELEC2} & Full CP (oracle) & .897 & 0.521 \\
 & CP+LS & .870 & 0.594 \\
 & nexCP+LS & .891 & 0.606 \\
 & nexCP+WLS & .893 & \textbf{0.509} \\
 & QT+WLS & .899 & 0.515 \\
\midrule
\multirow{5}{*}{Energy} & Full CP (oracle) & .915 & 280.5 \\
 & CP+LS & .920 & 355.8 \\
 & nexCP+LS & .906 & 308.7 \\
 & nexCP+WLS & .911 & \textbf{286.4} \\
 & QT+WLS & .929 & 364.1 \\
\midrule
\multirow{5}{*}{Bike} & Full CP (oracle) & .867 & 3192 \\
 & CP+LS & .740 & 3640 \\
 & nexCP+LS & .836 & 4078 \\
 & nexCP+WLS & .862 & 3190 \\
 & QT+WLS & .773 & \textbf{2512} \\
\midrule
\multirow{5}{*}{PRSA} & Full CP (oracle) & .915 & 148.1 \\
 & CP+LS & .883 & 191.4 \\
 & nexCP+LS & .899 & 195.9 \\
 & nexCP+WLS & .900 & 148.2 \\
 & QT+WLS & .894 & \textbf{148.1} \\
\bottomrule
\end{tabular}
\end{table}

Table~\ref{tab:results} reports the average coverage and width of each method, and Figure~\ref{fig:cov-width} shows their rolling counterparts. 
Additional results can be found in Appendices \cref{subsec:extresults} on \cref{fig:cov-width-ext}.

\paragraph{Synthetic data}
On the i.i.d. dataset, all online methods reach the nominal coverage ($0.894$ to $0.897$) with widths within $2\%$ of one another, so forgetting has no visible cost when the distribution is stable. 
The oracle slightly over-covers ($0.912$) with intervals $3$ to $4\%$ wider, consistent with the conservativeness of a finite-sample quantile computed on an effective sample of about $100$ points.
Under drift, the methods differ according to whether the base model forgets. With a static model, CP+LS under-covers ($0.844$ after the changepoints, $0.836$ under gradual drift), and nexCP+LS recovers $0.891$ and $0.886$ only by inflating its intervals by $72\%$ and $25\%$ with respect to the oracle ($7.13$ against $4.15$, $4.30$ against $3.45$): the quantile must absorb an error that it did not cause.
nexCP+WLS, whose model forgets, matches the oracle in both settings (coverages $0.903$ and $0.898$, widths $4.20$ and $3.36$). QT+WLS is as accurate under gradual drift ($0.892$, $3.30$) but under-covers by about $2.5$ points after abrupt breaks ($0.876$), as a tracker moving by increments of size $\eta$ needs several steps to recover.

\paragraph{Real world datastreams}
The four real streams confirm this ranking under drifts of unknown nature and intensity. 
On ELEC2 and PRSA, nexCP+WLS is almost indistinguishable from the oracle ($0.893$/$0.509$ against $0.897$/$0.521$, and $0.900$/$148.2$ against $0.915$/$148.1$), and so is QT+WLS ($0.899$/$0.515$ and $0.894$/$148.1$), whereas the static-model methods pay for the drift: nexCP+LS needs intervals $16\%$ and $32\%$ wider than the oracle's, and CP+LS under-covers ($0.870$ and $0.883$) despite intervals $14\%$ and $29\%$ wider.
On Energy, whose target is heavy-tailed, every method over-covers; nexCP+WLS again follows the oracle, while CP+LS, nexCP+LS and QT+WLS need intervals $27\%$, $10\%$ and $30\%$ wider.
Bike is the hardest stream, with only $731$ daily points and a strong trend in demand: even the oracle under-covers ($0.867$), nexCP+WLS matches it ($0.862$, width $3\,190$ against $3\,192$), CP+LS falls to $0.740$, and QT+WLS to $0.773$ with intervals $21\%$ narrower, as fewer than $600$ online steps leave the tracker too little time to follow the drift.

Overall, what separates the methods is whether the base model forgets, not how the quantile is computed.

\subsection{About the computational cost}
\label{subsec:comp}

The complexity of each method is summarized in \Cref{tab:complexity} and we provide, in \Cref{tab:time}, the average computational time of each method per step.

Storing residuals makes CP+LS and nexCP+LS $3$ to $8$ times slower than QT+WLS, yet with a static model this extra cost buys wider intervals, not better coverage. nexCP+WLS reaches the oracle's accuracy on every stream at three orders of magnitude less, and QT+WLS, with constant time and memory, matches it on smooth drifts, at the price of a slower reaction to abrupt changes.

\begin{table}[t]
\centering
\caption{Mean computation time per iteration (ms). Synthetic streams: mean
over 20 realizations. The oracle is averaged over its stride 20 evaluations.}
\label{tab:time}
\resizebox{\linewidth}{!}{%
\begin{tabular}{lccccccc}
\toprule
Method & i.i.d. & Changep. & Drift & ELEC2 & Energy & Bike & PRSA \\
\midrule
Full CP & 43.17 & 43.30 & 42.59 & 148.67 & 146.96 & 4.71 & 142.86 \\
CP+LS & 0.058 & 0.058 & 0.057 & 0.062 & 0.067 & 0.079 & 0.053 \\
nexCP+LS & 0.045 & 0.047 & 0.045 & 0.065 & 0.074 & 0.038 & 0.061 \\
nexCP+WLS & 0.047 & 0.052 & 0.052 & 0.069 & 0.088 & 0.032 & 0.057 \\
QT+WLS & \textbf{0.009} & \textbf{0.010} & \textbf{0.011} & \textbf{0.010} & \textbf{0.012} & \textbf{0.012} & \textbf{0.008} \\
\bottomrule
\end{tabular}
}
\end{table}

\section{Conclusion}
\label{sec:conclusion}

In this paper, we evaluated whether online conformal methods that adapt both the base model and the conformal quantile via forgetting factors can match the coverage and width of a true, unbounded history full CP oracle under distribution shift, rather than merely another online approximation. 
Across three synthetic regression regimes and four real-world benchmarks, which span different drift mechanisms, the methods that adapt the model (nexCP+WLS and QT+WLS) track the oracle's coverage and interval width closely, while the methods that adapt only the quantile and leave the model static (CP+LS and nexCP+LS) fall short under drift.  
Furthermore, the experiments illustrate that QT+WLS has memory and per-step cost of $O(p^2)$, independent of $t$, which makes it the only method suited to settings where storing the full history is not feasible.

\newpage

\appendix

\section{Scope of the Model-Update Comparison}

\label{remark:no_wls_analogue}

Concerning Section \ref{subsec:model_updates}, two gaps are worth stating explicitly rather than leaving implicit. 
First, QT+WLS has no exact full-CP counterpart: because it never revisits past residuals, there is no finite-history object it is a recursive approximation \emph{of}: its validity follows from the ACI guarantee, not from matching a full-CP limit, so we report it for reference in \Cref{subsec:res} but exclude it from the oracle-matching claim itself. 

\section{Experimental Protocol}
\label{subsec:protocol}

We consider seven regression streams: three synthetic settings and four real-world datasets.
In the three synthetic settings, $X_t\sim\mathcal N(0,I_4)$ and $Y_t=X_t^\top\beta_t+\varepsilon_t$ with $\varepsilon_t\sim\mathcal N(0,1)$, for $t=1,\dots,N=2000$ and without intercept. The covariates and the noise are shared by the three settings, which differ only through the coefficients $\beta_t$. 
In the \textit{i.i.d.} setting (setting 1), $\beta_t=(2,1,0,0)^\top$ is constant, which provides a stationary control.
In the changepoint setting (setting 2), $\beta_t$ successively equals $(2,1,0,0)^\top$, $(0,-2,-1,0)^\top$ and $(0,0,2,1)^\top$ on $[1,500)$, $[500,1500)$ and $[1500,2000]$. 
In the gradual drift setting (setting 3), $\beta_t$ moves linearly from $\beta_{\rm start}=(2,1,0,0)^\top$ to $\beta_{\rm end}=(0,0,2,1)^\top$, that is $\beta_t=\beta_{\rm start}+\frac{t}{N-1}(\beta_{\rm end}-\beta_{\rm start})$, so that the last two settings connect the same two extreme models through an
abrupt and a slow transition, respectively. 

The three synthetic streams are complemented by four real streams, whose drift mechanism, unlike that of the synthetic ones, is unknown.
ELEC2 \citep{harries1999splice} contains half-hourly records of the New South Wales electricity market. Following \citet{zaffran2022adaptive}, we discard the first $17\,760$ observations, during which the response is constant, and keep only the half-hours between 9:00 and 12:00, which leaves $N=3\,444$ points.
Energy \citep{candanedo2017data} records the consumption of household appliances every $10$ minutes, together with indoor and outdoor temperature and humidity measurements and two random variables, from 11 January 2016.
Bike \citep{fanaee2014event} gives the daily number of rented bicycles of the Capital Bikeshare system in 2011--2012 ($N=731$), with calendar and weather covariates.
PRSA \citep{liang2015assessing} reports the hourly PM2.5 concentration in Beijing with meteorological covariates, starting in January 2010; the hours with a missing response are removed.
For Energy, Bike and PRSA, the targets are, respectively, the appliances consumption, the daily count of rentals and the PM2.5 concentration. We remove the row identifiers and the covariates that are duplicated, constant or null on the training window, which would make the initial least-squares problem singular, as well as the casual and registered counts of Bike, whose sum is the response.
To keep the exact oracle tractable, Energy and PRSA are truncated to their first $N=3\,444$ observations, the length of ELEC2, which leaves $p=26$, $10$ and $11$ covariates for Energy, Bike and PRSA, against $p=4$ for ELEC2.
Finally, as the step size $\eta$ of QT+WLS is expressed in the unit of the response, it is set to $\eta=0.01\,\hat\sigma_Y$ on these three streams, where $\hat\sigma_Y$ is the standard deviation of the response on the training window; this is equivalent to standardizing the response and leaves the other methods unchanged.

The oracle is the exact weighted full-CP set of \cref{subsec:full_cp}, computed by the homotopy solver with the same exponential weights as nexCP+WLS, so that both solve the same weighted problem and differ only in whether the solution is recomputed from scratch or updated recursively. 
Because its cost grows with the history size, it is evaluated every $20$ steps; its coverage, width and run time are averaged over these evaluation times, which does not bias the averages but increases their variance. 
We report the average empirical coverage, the average width $\frac{1}{T-T_0}\sum_{t=T_0}^{T}2\hat q_t$, and the mean wall-clock time per iteration computing:

\begin{align*}
     \frac{1}{T-T_0}\sum_{t=T_0}^{T}\mathbf{1}\{Y_t\in\hat C_t(X_t)\},\; \text{and} \; 
    \frac{1}{T-T_0}\sum_{t=T_0}^{T}2\hat q_t
\end{align*} 

For the online methods this time covers the prediction, the quantile computation and the model update. 
The curves of Figure~\ref{fig:cov-width} are smoothed by a rolling average of
$50$ steps (synthetic) or $300$ steps (ELEC2).

\section{Additional Experimental Results}
\label{subsec:extresults}

\begin{figure}
    \centering 
    \includegraphics[width=0.48\linewidth]{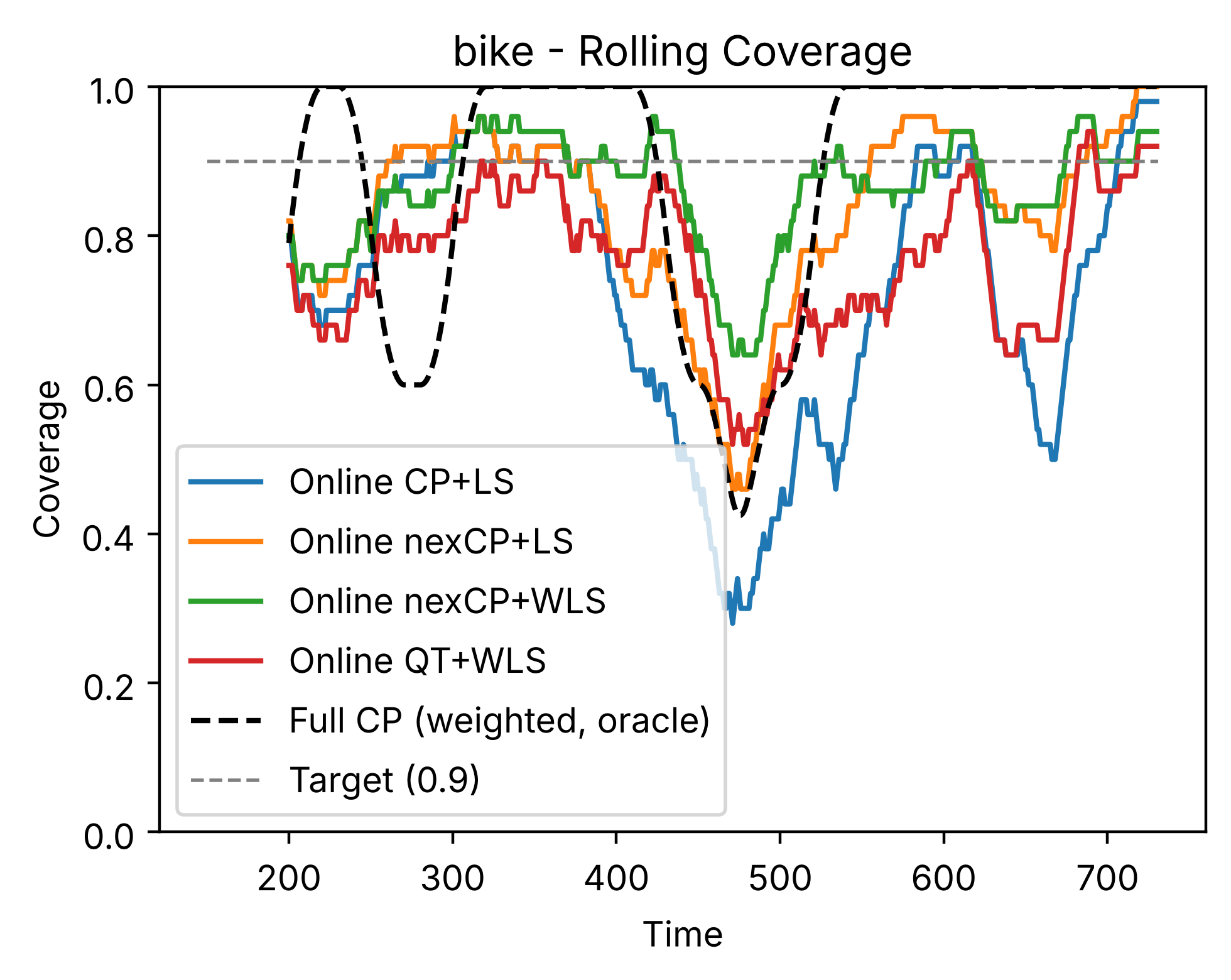} \includegraphics[width=0.48\linewidth]{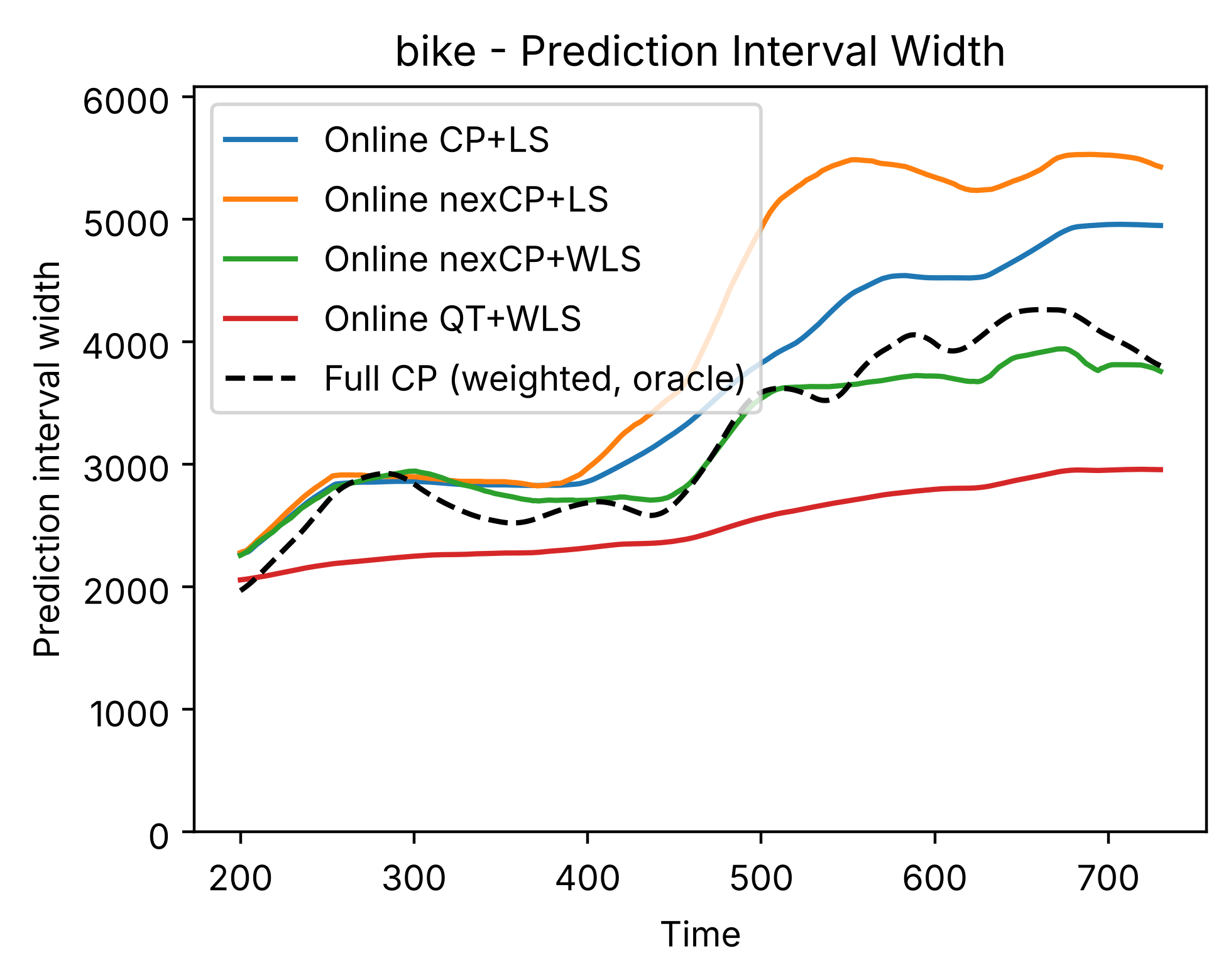}
    \includegraphics[width=0.48\linewidth]{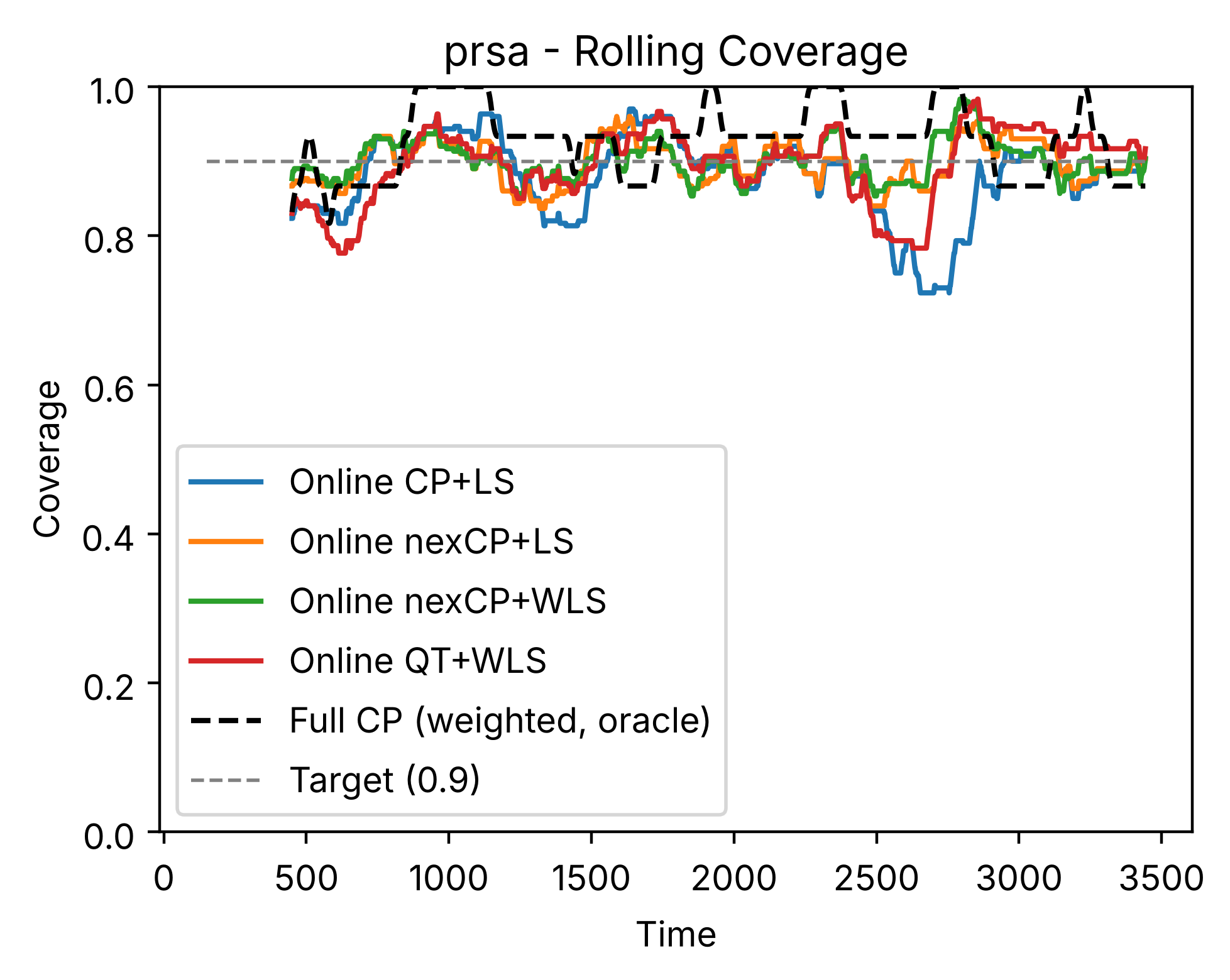} \includegraphics[width=0.48\linewidth]{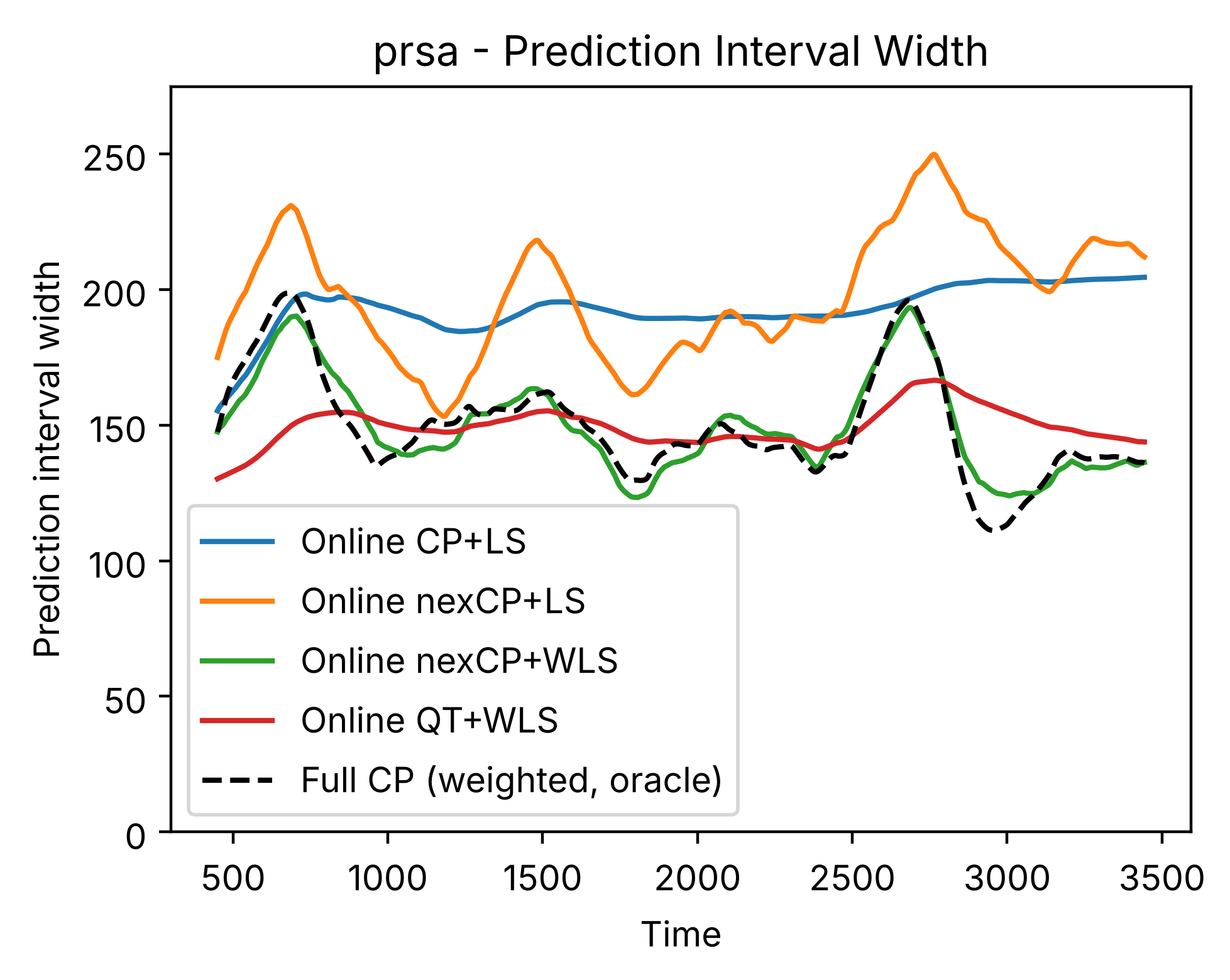}
    \includegraphics[width=0.48\linewidth]{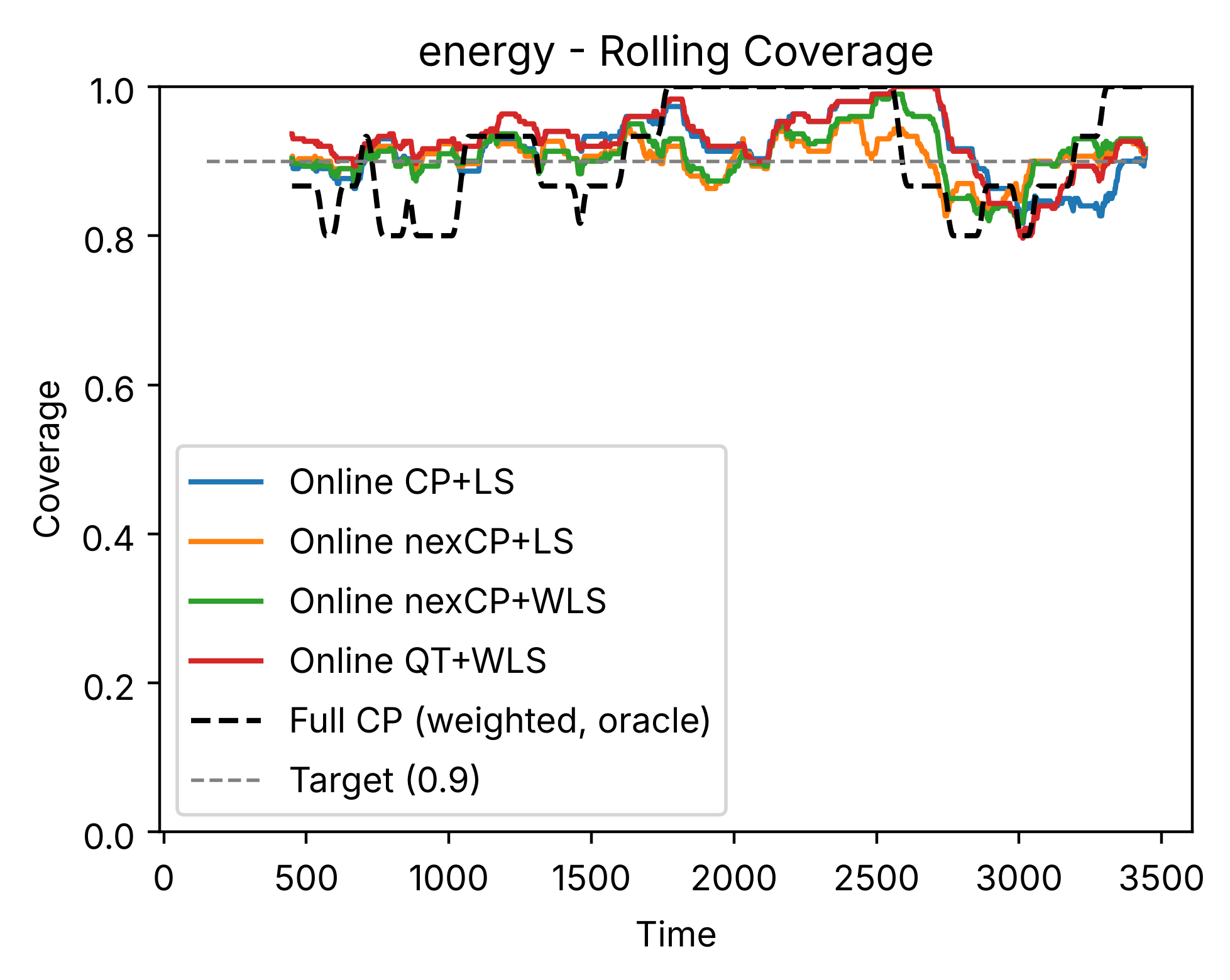} \includegraphics[width=0.48\linewidth]{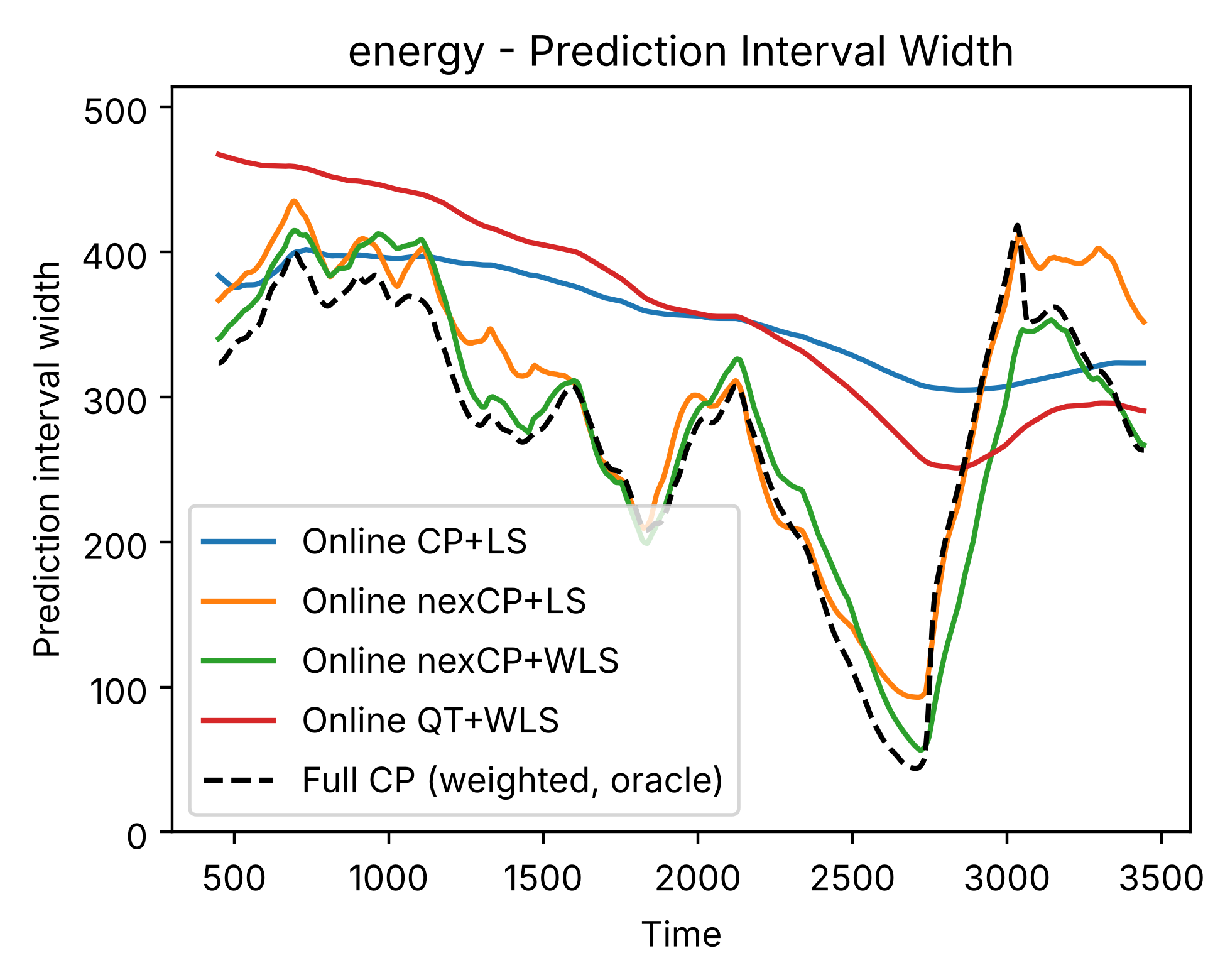}
    \caption{Rolling coverage (left) and interval width (right), online methods vs. the full-CP oracle on the Bike, PRSA and Energy datasets.}
    \label{fig:cov-width-ext}
\end{figure}

We present, on \cref{fig:cov-width-ext} the detailed results on the three remaining real data: Bike, PRSA and Energy, presenting the evolution of both coverage and interval width along time.

\clearpage
\bibliographystyle{abbrvnat}
\bibliography{references}

@article{angelopoulos2023conformal,
  title={Conformal prediction: A gentle introduction},
  author={Angelopoulos, Anastasios N and Bates, Stephen},
  journal={Foundations and Trends in Machine Learning},
  volume={16},
  number={4},
  pages={494--591},
  year={2023},
  publisher={Emerald Publishing Limited}
}

@article{angelopoulos2023pid,
  title={Conformal pid control for time series prediction},
  author={Angelopoulos, Anastasios and Candes, Emmanuel and Tibshirani, Ryan J},
  journal={Advances in neural information processing systems},
  volume={36},
  pages={23047--23074},
  year={2023}
}

@article{angelopoulos2024online,
  title={Online conformal prediction with decaying step sizes},
  author={Angelopoulos, Anastasios N and Barber, Rina Foygel and Bates, Stephen},
  journal={arXiv preprint arXiv:2402.01139},
  year={2024}
}

@inproceedings{bhatnagar2023improved,
  title={Improved online conformal prediction via strongly adaptive online learning},
  author={Bhatnagar, Aadyot and Wang, Huan and Xiong, Caiming and Bai, Yu},
  booktitle={International Conference on Machine Learning},
  pages={2337--2363},
  year={2023},
  organization={PMLR}
}

@article{barber2023conformal,
  title={Conformal prediction beyond exchangeability},
  author={Barber, Rina Foygel and Candes, Emmanuel J and Ramdas, Aaditya and Tibshirani, Ryan J},
  journal={The Annals of Statistics},
  volume={51},
  number={2},
  pages={816--845},
  year={2023},
  publisher={Institute of Mathematical Statistics}
}

@article{gibbs2021adaptive,
  title={Adaptive conformal inference under distribution shift},
  author={Gibbs, Isaac and Candes, Emmanuel},
  journal={Advances in Neural Information Processing Systems},
  volume={34},
  pages={1660--1672},
  year={2021}
}

@article{gibbs2024conformal,
  title={Conformal inference for online prediction with arbitrary distribution shifts},
  author={Gibbs, Isaac and Cand{\`e}s, Emmanuel J},
  journal={Journal of Machine Learning Research},
  volume={25},
  number={162},
  pages={1--36},
  year={2024}
}

@article{harries1999splice,
  title={Splice-2 comparative evaluation: Electricity pricing},
  author={Harries, Michael and Wales, New South and others},
  year={1999},
  publisher={University of New South Wales, School of Computer Science and Engineering~…}
}

@article{lei2018distribution,
  title={Distribution-free predictive inference for regression},
  author={Lei, Jing and G’Sell, Max and Rinaldo, Alessandro and Tibshirani, Ryan J and Wasserman, Larry},
  journal={Journal of the American Statistical Association},
  volume={113},
  number={523},
  pages={1094--1111},
  year={2018},
  publisher={Taylor \& Francis}
}

@article{tibshirani2019conformal,
  title={Conformal prediction under covariate shift},
  author={Tibshirani, Ryan J and Foygel Barber, Rina and Candes, Emmanuel and Ramdas, Aaditya},
  journal={Advances in neural information processing systems},
  volume={32},
  year={2019}
}

@inproceedings{vovk2002online,
  title={On-line confidence machines are well-calibrated},
  author={Vovk, Vladimir},
  booktitle={The 43rd Annual IEEE Symposium on Foundations of Computer Science, 2002. Proceedings.},
  pages={187--196},
  year={2002},
  organization={IEEE}
}

@book{vovk2005algorithmic,
  title={Algorithmic learning in a random world},
  author={Vovk, Vladimir and Gammerman, Alexander and Shafer, Glenn},
  year={2005},
  publisher={Springer}
}

@inproceedings{zaffran2022adaptive,
  title={Adaptive conformal predictions for time series},
  author={Zaffran, Margaux, et al},
  booktitle={International conference on machine learning},
  pages={25834--25866},
  year={2022},
  organization={PMLR}
}

@article{candanedo2017data,
  title   = {Data driven prediction models of energy use of appliances in a low-energy house},
  author  = {Candanedo, Luis M. and Feldheim, V{\'e}ronique and Deramaix, Dominique},
  journal = {Energy and Buildings},
  volume  = {140},
  pages   = {81--97},
  year    = {2017}
}

@article{fanaee2014event,
  title   = {Event labeling combining ensemble detectors and background knowledge},
  author  = {Fanaee-T, Hadi and Gama, Jo{\~a}o},
  journal = {Progress in Artificial Intelligence},
  volume  = {2},
  number  = {2},
  pages   = {113--127},
  year    = {2014}
}

@article{liang2015assessing,
  title   = {Assessing {B}eijing's {PM}2.5 pollution: severity, weather impact, {APEC} and winter heating},
  author  = {Liang, Xuan and Zou, Tao and Guo, Bin and Li, Shuo and Zhang, Haozhe and Zhang, Shuyi and Huang, Hui and Chen, Song Xi},
  journal = {Proceedings of the Royal Society A},
  volume  = {471},
  number  = {2182},
  pages   = {20150257},
  year    = {2015}
}

\end{document}